\documentclass[twoside,final]{pro12}

\usepackage{amsmath}
\usepackage{tabularx,multirow,hhline}
\usepackage{graphicx}
\usepackage{textcomp}

\head{L. Lamy et al.} {40 years of decametric observations with the NDA}	

\begin{document}

\title{1977--2017: 40 YEARS OF DECAMETRIC OBSERVATIONS OF JUPITER AND THE SUN WITH THE NANCAY DECAMETER ARRAY}	
\author{L. Lamy\adress{\textsl LESIA, Observatoire de Paris, CNRS, PSL, UPMC, UPD, Meudon, France}$\,$, P. Zarka$^*$, B. Cecconi$^*$, L. Klein$^*$,\\ S. Masson$^*$, L. Denis\adress{\textsl USN, Observatoire de Paris, CNRS, PSL, UO/OSUC, Nan\c cay, France}$\,$, A. Coffre$^\dag$, and C. Viou$^\dag$}

\maketitle

\begin{abstract}
The Nan\c cay Decameter Array (NDA) routinely observes low frequency (10--100~MHz) radio emissions of Jupiter and the Sun since 4 decades. The NDA observations, acquired with a variety of receivers with increasing performances, were the basis for numerous studies of jovian and solar radio emissions and now form a unique long-term database spanning $\ge 3$ solar cycles and jovian revolutions. In addition, the NDA historically brought a fruitful support to space-based radio observatories of the heliosphere, to multi-wavelength analyses of solar activity and contributes to the development of space weather services. After having summarized the NDA characteristics, this article presents latest instrumental and database developments, some recent scientific results and perspectives for the next decade.\end{abstract}

\section{The Nan\c cay Decameter Array, a unique facility}

The Nan\c cay Decameter Array (NDA) has been built between 1975 and 1977 within the radio astronomy station near the village of Nan\c cay (Sologne, France) to routinely observe jovian and solar decametric (DAM) emissions in the 10--100~MHz (30--3~m) range from the ground. Early decametric observations were previously acquired in Nan\c cay with two log-periodic tracking antennae since 1970 [Boischot, 1974]. The NDA characteristics are fully detailed in [Boischot et al., 1980; Lecacheux, 2000] and only briefly reminded below.\\

The NDA is a high gain, phased, array of relatively modest size. It consists of 144 helical antennae, which corresponds to a $\sim$7000~m$^2$ effective area at 25~MHz. It is composed of 2 sub-arrays, right-handed and left-handed polarized, with a wide main lobe ($\sim6\times10^\circ$). Periodic (hourly) calibration sequences enable one to derive absolute flux densities.

The full NDA observes Jupiter and the Sun on a daily basis (at best $\sim$8~h per day each) since January 1978. The observations were acquired with a series of analog ($\le$1990) and digital ($\ge$1990) receivers. These observations now form a unique long-term database which cover more than 3 solar cycles and 3 jovian revolutions. Their analysis were the basis of numerous studies of jovian and solar radio emissions. 

The NDA additionally brought a fruitful support to space-based radio observatories of Jupiter (such as Voyager, Galileo and now Juno) and the Sun (Wind, Stereo and soon Solar Orbiter/Solar Probe+), to multi-wavelength analyses of solar activity (through Radio Monitoring or BASS2000 databases) and contributes to the development of space weather services (for instance with the project FEDOME). In this article, we briefly describe the available types of NDA measurements and the current data distribution in section \ref{data}. We then present some recent results in section \ref{results} before turning to perspectives for the next decade.

\section{A unique dataset}
\label{data}

\subsection{A variety of receivers}

\begin{figure}[ht]
\centering
\includegraphics[width=0.9\textwidth]{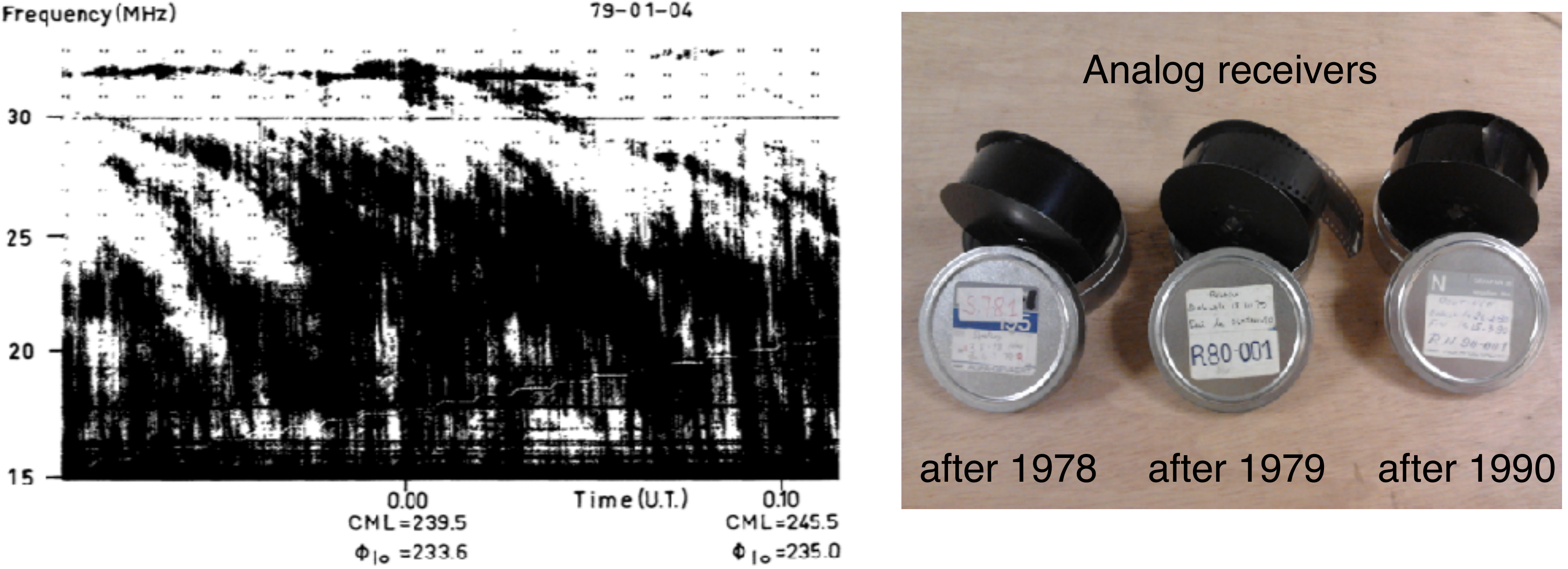}
\caption{(left) NDA analog observations of Jupiter DAM in 1979 directly printed on a fac simile device connected to a spectrum analyzer (SFA), just prior to the Voyager flyby [Genova et al., 1981]. (Right) Picture of archive 35-mm films recorded with 3 analog receivers.}
\label{fig1}
\end{figure}

From 1975 (construction of the NDA initial set of 32 antennae) to 1990, the observations were acquired mainly with 3 different analog receivers. The data were recorded on 35-mm films and simultaneously drawn on fac-similes (Figure \ref{fig1}). This unique dataset, inventoried in 2017, is worth for digitization but its huge volume (900 films, which correspond to a total integrated length of about 27~km !) together with a variety of instrumental modes and the lack of documentation is challenging.

\begin{figure}[ht]
\centering
\includegraphics[width=0.8\textwidth]{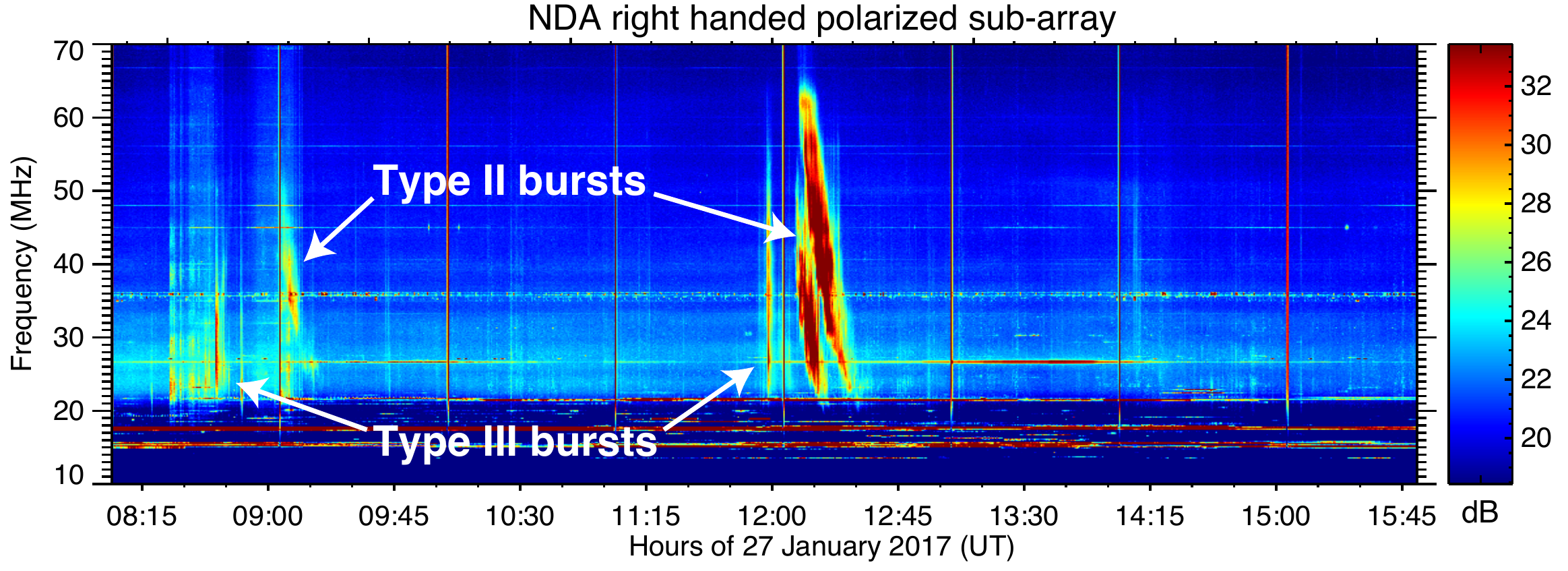}
\caption{NDA observations of solar radio bursts on 27 January 2011 [Pick et al., 2016] with the Routine digital receiver.}
\label{fig2}
\end{figure}

\begin{figure}[ht]
\centering
\includegraphics[width=\textwidth]{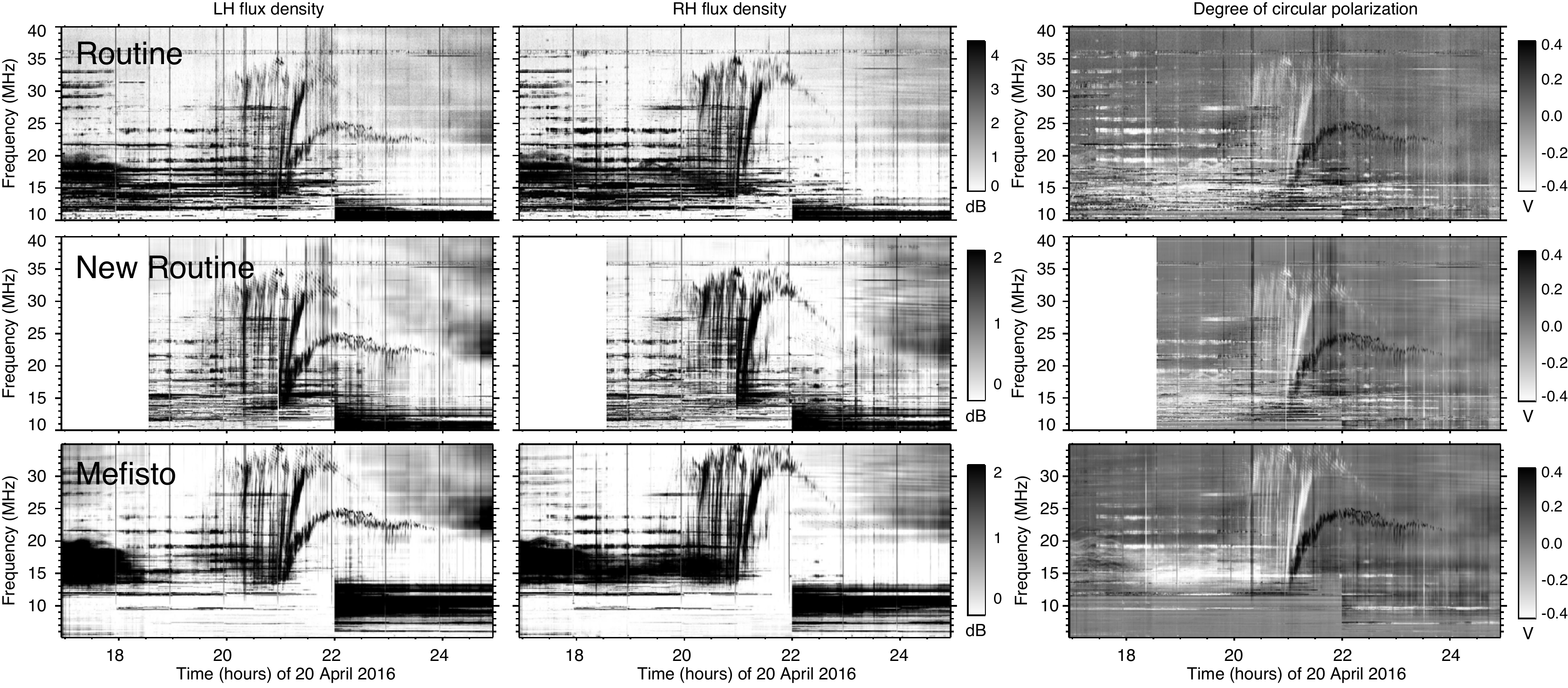}
\caption{NDA observations of Io-B (RH polarized, northern hemisphere) and Io-D (LF polarized, southern hemisphere) DAM emissions acquired simultaneously with the Routine, New-Routine and Mefisto receivers on 20 April 2016. The New-Routine measurements are more sensitive than the Routine ones. The Mefisto data display less intense RFI, owing to the median filtering.}
\label{fig3}
\end{figure}

\begin{figure}[ht]
\centering
\includegraphics[width=\textwidth]{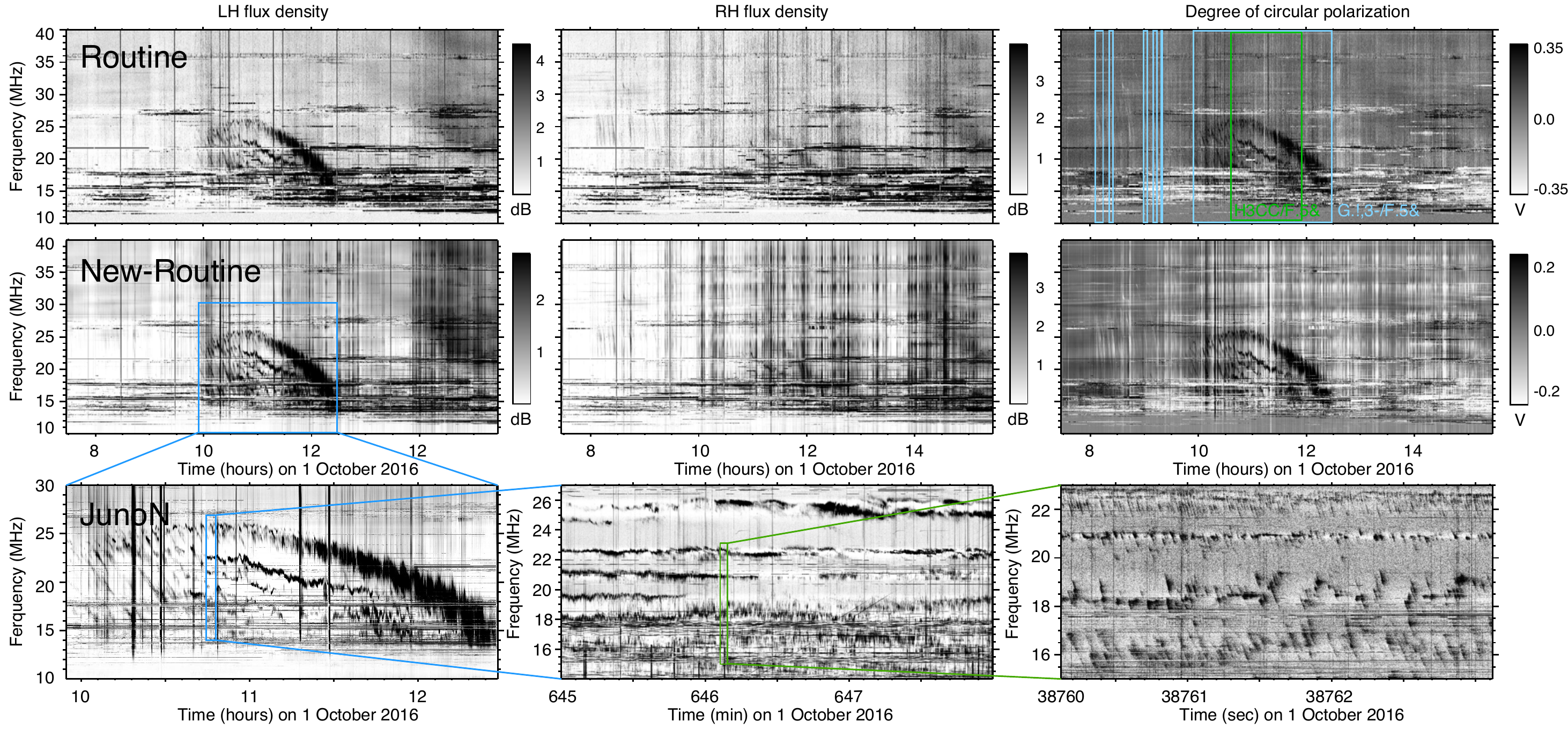}
\caption{NDA observations of Io-C (LH polarized, southern hemisphere) emission acquired simultaneously with the Routine, New-Routine and JunoN receivers on 1 October 2016. The New-Routine observations are more sensitive than the Routine ones. The blue boxes (green resp.) on the top right panel display intervals when JunoN observations were kept at MR (FR resp.) resolution. The bottom panel displays a series of zooms of JunoN measurements of Io-C DAM with a total duration of 2.5~h, 3~min and 3~s from left to right, respectively. The bottom right panel clearly reveals S-bursts.}
\label{fig4}
\end{figure}

Since 1990, numerous digital receivers have been developed in Nan\c cay and connected to the decameter array, enabling a variety of measurements including the determination of part of or all Stokes parameters, increasing time--frequency resolutions up to waveform measurements [Zarka, 2007]. Table 1 lists the 4 digital receivers still in operation. The historical Routine receiver measures LH and RH flux densities in alternance at modest t--f resolution (see examples in Figures \ref{fig2}-\ref{fig4}). 

Since 2012, 3 new receivers were successively connected to the array. The versatile New Routine receiver [Zarka et al., 2012] measures instantaneous auto- and cross-correlations between the LH and RH sub-arrays at high t--f resolution (see Figures \ref{fig3}-\ref{fig4}). The temporal resolution is set up to 500~ms by default but can be increased up to 1~ms. The Mefisto receiver [Lecacheux et al., 2013] is aimed at recovering the LF part of the spectrum by applying on--the--fly median filtering. It measures simultaneously LH and RH flux densities at high t--f resolution but limits to the 10--35~MHz range (see Figure \ref{fig3}). 

Finally, the JunoN (for Juno--Nan\c cay) receiver was installed in early 2016, to support the Juno mission. This analyser is based on a Digital Down Conversion (DDC) able to select a 50-MHz complex band within the 0-100 MHz range. Particular care was taken to provide precise time stamping with respect to the local GPS/Rb-driven reference clock. The raw data stream can be recorded directly (waveform mode) or channelized with a 16k-FFT tapered with a Blackman-Harris function (dynamic spectrum mode). JunoN measures auto- and cross-correlations between the LH and RH sub-arrays at very high t--f resolution. Two standard resolutions are used: full resolution (FR) at 2.6~ms$\times$3.05~kHz and medium resolution (MR) at 83.2~ms$\times$12.2~kHz. Due to the enormous volume of produced data, JunoN MR measurements are kept only when Jovian radio emissions are detected and FR measurements whenever fine structures are observed. For the second half of 2016, JunoN data typically reached a few Tb per day. An example of JunoN observations is displayed in Figure \ref{fig4}.

\begin{table}[h]
\caption{NDA digital receivers in operation. The measurements notations RR, LL and LR respectively indicate autocorrelation on the RH array, autocorrelation on the LH array, and complex cross-correlation between both arrays. Notes : (1) RR and LL are measured alternatively ; (2) both filtered and unfiltered measurements are available; (3) full resolution (4) medium resolution.}\label{Tab.3}
\scriptsize{
\begin{center}
\begin{tabularx}{\linewidth}{|p{1.8cm}|p{2cm}|p{1.1cm}|p{2.3cm}|p{2cm}|p{1.7cm}|X|} 
\hline 
Receiver & Measurements & Channels & Spectral range& Resolution& Data volume& Format \\
 & & & (MHz) & (ms~$\times$~kHz)& (per 8~h) &  \\
\hline
\multirow{2}{1.8cm}{Routine\\[.2ex] (1990--...)}&\multirow{2}{2cm}{RR, LL$^1$}& \multirow{2}{2.3cm}{2} & 10--40  (Jupiter) & 500 $\times$ 75~kHz & \multirow{2}{*}{22~MB / 8h}& binary, CDF\\ \hhline{~~~--~-}
 & & & 10--80  (Sun) & 500 $\times$ 175 & & binary, CDF\\
\hline
\multirow{2}{1.8cm}{New Routine\\[.2ex] (2012--...)}&\multirow{2}{2cm}{RR, LL, LR}& \multirow{2}{2cm}{4}  & 10--40 (Jupiter) & \multirow{2}{2.9cm}{500 $\times$ 49}& 700~MB& binary, FITS\\ \hhline{~~~-~--}
& & & 10--88  (Sun) & &1.8~GB& binary, FITS \\
\hline
Mefisto (2013--...)& \multirow{2}{2cm}{RR, LL}&\multirow{2}{2cm}{2+2$^2$}&\multirow{2}{2cm}{10--35}&\multirow{2}{2cm}{100 $\times$ 80}&\multirow{2}{2cm}{2.6~GB}& \multirow{2}{2cm}{binary, FITS}\\
\hline
\multirow{2}{1.8cm}{JunoN\\[.2ex] (2016--...)}& \multirow{2}{2cm}{RR, LL, LR + Waveform}&\multirow{2}{*}{4} & \multirow{2}{2.3cm}{6--56  (Jupiter)}& 2.6 $\times$ 3.05$^3$& 2.9~TB& \multirow{2}{1.7cm}{binary}\\\hhline{~~~~--~}
& & & & 83.2 $\times$ 12.2$^4$&22.6~GB & \\
\hline
\end{tabularx}
\end{center}
}
\end{table}

\subsection{Data distribution}

\begin{figure}[ht]
\centering
\includegraphics[width=\textwidth]{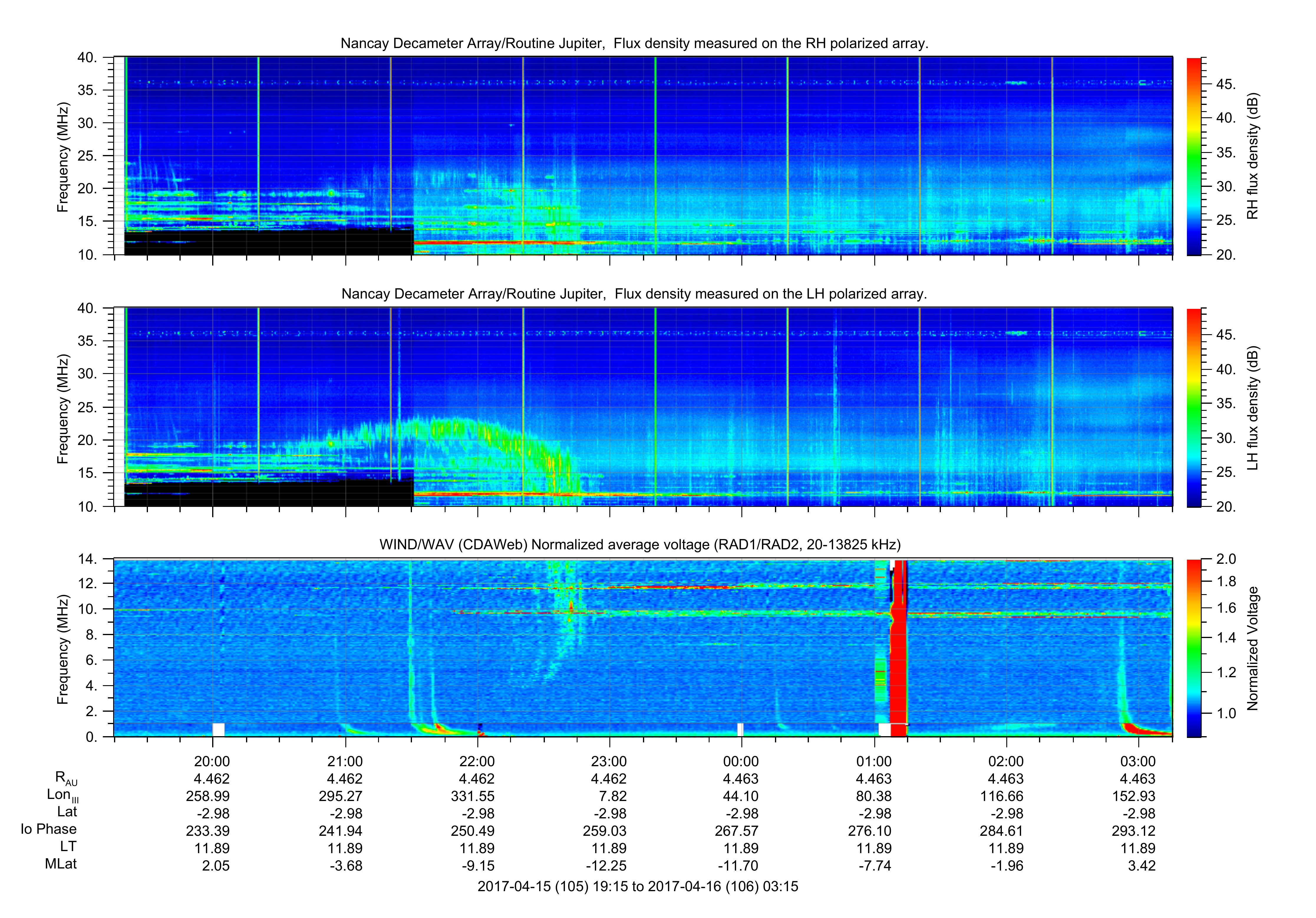}
\caption{Visualization of NDA and Wind/Waves observations of Jupiter across the full Jupiter spectrum. The Autoplot tool is able to read and display interactively NDA CDF online data ahead of Wind/Waves CDAWeb data.}
\label{fig5}
\end{figure}

The diverse NDA data described above is currently distributed as follows. All the public data are made available to the community on the NDA webpage \footnote{https://www.obs-nancay.fr/-Le-reseau-decametrique-.html}. Routine raw data are released every day, and were originally provided in a binary format. Since early 2017, Jupiter Routine raw data are also provided in the VO-compliant CDF format \textbf{[Cecconi et al., 2016]}. The Jupiter Routine CDF database cant thus be queried by VO portals, such as VESPA [Erard et al., 2017] and/or can be plotted interactively with VO visualization tools such as Autoplot [Faden et al., 2010]. An example of NDA Routine observations of Jupiter plotted ahead of those of Wind/Waves, both covering the full Jupiter spectrum, is displayed in Figure \ref{fig5}. New Routine and Mefisto raw data are released 1~year after their acquisition, in standard format (FITS and/or CDF). JunoN data are not released publicly yet but can be queried on demand.

\section{Recent scientific results}
\label{results}

Scientific results based on NDA observations spanning the range from 1978 to 2000 (for the solar Corona) and to 2007 (for Jupiter) have been reviewed by [Boischot et al., 1980; Lecacheux, 2000; Zarka, 2007]. Some examples of more recent studies are indicated below.

\subsection{The solar corona}


The NDA is the most sensitive whole--Sun routine spectrograph in the 20--80~MHz range. Bursty radio emission at these frequencies comes from altitudes between roughly 0.5 and 1 solar radius above the photosphere, a key region for solar-heliospheric connection studies. Work on solar radio bursts published since 2003 addressed the following topics.

\subsubsection{Propagation of electron beams through the high corona} Coronal electrons beams are known to produce type III radio emissions drifting in (generally toward low) frequency [e.g. Melnik et al., 2015] which, in turn, are a tracer of field lines that connect particle acceleration regions in solar active regions or coronal shock waves to the interplanetary space. NDA observations have been used to identify closed and open magnetic field configurations in the solar corona, notably in studies of the origin and propagation of solar energetic particles detected near the Earth [Kerdraon et al., 2010; Klein et al., 2010; Malandraki et al., 2012].

At times not specifically related to flare occurrence or to the solar cycle, groups of faint frequency drifting emissions (similar to Òsolar S burstsÓ) were detected with the NDA and the UTR-2 radiotelescopes. Statistical study of their drift rates (df/dt) and durations ($\Delta$t) allowed Briand et al. [2008] to identify three populations, with respectively df/dt $\sim$ -1, 0 and +1 MHz/s and $\Delta$t $\sim$  11, 1 and 3 seconds. These emissions are attributed to Òelectron cloudsÓ propagating through the solar corona at 3Ð5 times the electron thermal velocity. Vlasov-Ampre simulations suggested that localized, time-dependent electron heating in the corona naturally generates such low velocity electron clouds which, in turn, can generate Langmuir waves and electromagnetic signals by nonlinear processes.

At times of high solar activity, decameter spikes have also been observed and shown to be the low-frequency analogues of high-frequency (GHz) solar spikes and used to derive the coronal plasma parameters [Shevchuk et al., 2016].

\subsubsection{Formation and propagation of coronal shock waves} Type II radio emission drifting toward low frequencies is the oldest known tracer of shock waves in the corona, and still their most direct diagnostic. In several recent studies, the radio emission was used to relate shock waves to their parent processes in the corona and to specifically investigate whether Coronal Mass Ejections (CME) are the unique drivers of shocks [Magdalenic et al., 2012]. Key parameters of shocks such as speed and Mach number, as well as the magnetic field strength in the ambient corona, were inferred from detailed analyses of the radio spectrum [Zucca et al., 2014; Salas--Matamoros et al., 2016]. Type IV bursts are another type of intense wideband radio emissions generally associated with fast CMEs and solar X flares, whose properties have been statistically assessed by [Hillaris et al., 2016]. The high NDA sensitivity makes it a particularly well-suited instrument for this research field.

\subsubsection{Early development of CMEs and particle acceleration in the corona} CMEs are most clearly seen in white-light coronographic observations, but those do not show the processes at the liftoff and during the early propagation of the ejecta. In addition, they only reveal the thermal plasma. Non thermal processes, including those occurring at the launch of CMEs, can be studied with radio observations. NDA has been used to constrain the propagation of CMEs in their early phase, and to show that it is frequently associated with electron acceleration. Prolonged electron acceleration that extends well beyond the impulsive flare phase has been observed [Carley et al., 2016; Koval et al., 2016]. The timing and frequency range determined from NDA data could be employed to trace the acceleration processes back to the interaction between the CME front and the ambient corona [Pick et al., 2016; Salas--Matamoros et al., 2016] (Figure \ref{fig2} provides an illustration of the former study), e.g. through magnetic reconnection, and to current sheets that form behind the outward--propagating magnetic structure [Huang et al., 2011].

The above--described work relies on multi-instrument analyses. The NDA is playing an essential role in supporting the major projects of solar-heliospheric physics such as the SoHO and STEREO missions and other ground-based radio telescopes such as UTR-2, URAN, the VLA or LOFAR [Konovalenko et al., 2000; Konovalenko et al., 2016]. Besides its support for scientific research, NDA, when observing the Sun, also provides monitoring data for space weather services, especially the FEDOME experimental space weather centre of the French air force. 

\subsection{Jupiter}

Jupiter decametric (DAM) emission extends up to 40~MHz. It divides into radiation related to the main auroral oval and radiation driven by planet--satellite interactions with the Galilean satellites. Studies published since 2007 dealt with the analysis of Io and non-Io DAM emissions.

\subsubsection{Statistical analysis and solar wind control}

A catalog of 26 years of NDA observations of Jupiter was recently built by [Soarez--Marques et al., this issue]. This work enabled the authors to perform a statistical analysis of Io and non-Io DAM properties (occurrence probability as a function of central meridian longitude of the observer and Io phase, occurrence as a function of time etc.). For non-Io DAM events of the same catalog, a statistical occurrence study led [Zarka et al., this issue] to unambiguously identify emissions specifically controlled by Ganymede and their average properties. 

Hess et al. [2012] compared 3 years of NDA observations of Jupiter over 1996--1999 with solar wind propagated parameters to demonstrate that the occurrence and intensity of non-Io DAM emissions is controlled by the solar wind dynamic pressure. A subsequent multi-point study using NDA, Cassini and Galileo radio observations late 2000 (just before the Cassini flyby of Jupiter) further showed that interplanetary fast forward (reverse resp.) shocks trigger dusk (both dawn and dusk resp.) auroral radio sources, which then significantly sub-corotate with the planet at 50--80\% of the rotation period [Hess et al., 2014]. 

\subsubsection{Support to other radio observations}

Nigl et al. [2007] used the NDA to test early LOFAR performances with VLBI (700~km baseline) high resolution measurements of Io-DAM S-bursts. 

More recently, a coordinated effort of ground-based radio telescopes (including the NDA) to support the Juno mission resulted in the formation of the Juno Ground-based Radio group [Cecconi et al., this issue]. Coordinated observations have been acquired since early 2016, before the Juno arrival to Jupiter. [Lecacheux et al., this issue] analyzed comparatively high resolution measurements acquired with the NDA, the LWA and the URAN-2 facilities to study the linear polarization of non-Io DAM as seen through the terrestrial ionosphere. The Juno approach then provided an increasing number of opportunities to directly compare Juno/Waves observations [Kurth et al., this issue] to ground-based ones. [Imai et al., this issue] for instance studied two non-Io DAM arcs successively seen by Juno/Waves and the NDA to assess the beaming angle of the emission [Louis et al., 2017].

\section{Conclusion}

40 years after its completion, the NDA remains operational and continues the daily acquisition of high resolution jovian and solar observations with sophisticated digital receivers. The recorded observations now form a unique database, spanning $\ge 3$ solar cycles and jovian revolutions, enabling long-term statistical studies and/or multi-wavelength analyses of solar activity. Such studies shall be further facilitated by the ongoing re-organization of (VO-compliant) NDA data formats, levels and distribution. The NDA continues to bring an ideal ground-based support to space missions, such as Juno and, soon, Solar Orbiter, Solar Probe+ and then JUICE in synergy with other ground-based radiotelescopes such as UTR-2, URAN, the VLA or LOFAR [Konovalenko et al., this issue]. In parallel, the NDA contributes to the development of space weather services.

\section*{Acknowledgements}
The authors were supported by the PNP and PNST programs of CNRS/INSU. The NDA observations are accessible online at http://www.obs-nancay.fr. The NDA is hosted by the Nan\c cay Radio Observatory / Unit\'e Scientifique de Nan\c cay of the Observatoire de Paris (USR 704-CNRS, supported by Universit\'e d\textquotesingle Orl\'eans, OSUC, and R\'egion Centre in France).


\section*{References}
\everypar={\hangindent=1truecm \hangafter=1}


Boischot,~A., Radioastronomy on decameter wavelengths at Meudon and Nan\c cay observatories, \textsl{Sol. Phys.}, \textbf{36}, 517--522, 1974.

Boischot,~A., C.~Rosolen, M.\,G.~Aubier, G.~Daigne, F.~Genova, Y.~Leblanc, A.~Lecacheux, J.~de~la~No\"e, and B.\,M.~Pedersen, A New High Gain, Broadband Steerable, Array to Study Jovian Decametric Emission, \textsl{Icarus}, \textbf{43}, 399--407, 1980.

Briand,~C., A.~Zasvlasky, M.~Maksimovic, P.~Zarka, A.~Lecacheux, H.~O.~Rucker, A.~A~Konovalenko, E.~P.~Abranin, V.~V.~Dorovsky, A.~A.~Stanislavsky, V.~N.~Melnik, Faint solar structures from decametric radio observations, \textsl{Icarus}, \textbf{Astron. Astrophys.}, \textbf{490}, 339--344, 2008..

Carley,~E.\,P., N.~Vilmer, and P.\,T.~Gallagher, Radio Diagnostics of Electron Acceleration Sites During the Eruption of a Flux Rope in the Solar Corona, \textsl{Astrophys. J.}, \textbf{833}, 87, 2016.

Cecconi,~B., A.~Coffre, L.~Denis and L.~Lamy, SRN NDA Routine Jupiter EDR CDF Dataset Specification, doi:10.5281/zenodo.831505, 2016.

Erard,~S. and 41 coauthors, VESPA: A community-driven Virtual Observatory in Planetary Science, \textsl{Planet. and Sp. Sci.}, in press, 2017.

Faden,~J., R.~S.~Weigel, J.~Merka and R.~H.~W.~Friedel, Autoplot: a Browser for Scientific Data on the Web,  \textsl{Earth Sci. Inform.}, \textbf{3}, 41--49, 2010.

Genova,~F., M.\,G.~Aubier, and A.~Lecacheux, Modulations in {J}ovian decametric spectra: {P}ropagation effects in terrestrial ionosphere and {J}ovian environment, \textsl{Astron. Astrophys.}, \textbf{104}, 229--239, 1981.

Hess,~S.~L.~G., E.~Echer and P.~Zarka, Solar wind pressure effects on Jupiter decametric radio emissions independent of Io, \textsl{Planet. and Sp. Sci.}, \textbf{70}, 114--125, 2012.

Hess,~S.~L.~G., E.~Echer, P.~Zarka, L.~Lamy and P.~A.~Delamere, Multi-instrument study of the Jovian radio emissions triggered by solar wind shocks and inferred magnetospheric subcorotation rates, \textsl{Planet. and Sp. Sci.}, \textbf{99}, 136--148, 2014.

Hillaris,~A., C.~Bouratzis and A.~Nindos, Interplanetary Type IV Bursts, \textsl{Sol. Phys.}, \textbf{291}, 2049--2069, 2016.

Huang,~J., P.~D{\'e}moulin, M.~Pick, F.~Auch{\`e}re, Y.\,H.~Yan and A.~Bouteille, Initiation and Early Development of the 2008 April 26 Coronal Mass Ejection, \textsl{Astrophys. J.}, \textbf{729}, 107, 2011.

Kerdraon,~A., M.~Pick, S.~Hoang, Y.\,M.~Wang and D.~Haggerty, The Coronal and Heliospheric 2007 May 19 Event: Coronal Mass Ejection, Extreme Ultraviolet Imager Wave, Radio Bursts, and Energetic Electrons, \textsl{Astrophys. J.}, \textbf{715}, 468--476, 2010.

Klein,~K.--L., G.~Trottet and A.~Klassen, Energetic Particle Acceleration and Propagation in Strong CME-Less Flares, \textsl{Sol. Phys.}, \textbf{263}, 185--208, 2010.

Koval,~A., A.~Stanislavsky, Y.~Chen, S.~Feng, A.~Konovalenko and Y.~Volvach, A Decameter Stationary Type IV Burst in Imaging Observations on 2014 September 6, \textsl{Astrophys. J.}, \textbf{826}, 125, 2016.

Konovalenko,~A.~A., A.~Lecacheux and C.~Rosolen, Large Ground Based Array Antennas for Very Low Frequency Radio Astronomy, Proceedings of the Radio Astronomy at Long Wavelengths, Perspectives on Radio Astronomy: Technologies for Large Antenna Arrays conference held in Dwingeloo, Eds A. B. Smolders and M. P. Haarlem, published by ASTRON, 115, 2000.

Konovalenko,~A.~A. and 71 coauthors, The modern radio astronomy network in Ukraine: UTR-2, URAN and GURT,  \textsl{Exp. Astron.}, \textbf{42}, 11--48, 2016.

Louis,~C.,~K., L.~Lamy, P.~Zarka, B.~Cecconi, M.~Imai, W.~S.~Kurth, G.~Hospodarsky, S.~L.~G.~Hess, X.~Bonnin, S.~Bolton, J.~E.~P.~Connerney and S.~M.~Levin, Io--Jupiter decametric arcs observed by Juno/Waves compared to ExPRES simulations, \textsl{Geophys. Res. Lett.}, in press.

Lecacheux,~A., C.~Dumez--Viou and K.-L.~Klein, Un spectrographe pour la radioastronomie aux ondes courtes, au voisinage de la coupure ionospherique, Journees scientifiques URSI--France, Paris, 73--76, 2013.

Lecacheux,~A., The Nan\c{c}ay Decameter Array: A useful step towards giant, new generation radio telescopes for long wavelength radio astronomy, Radio Astronomy at Long Wavelengths, \textsl{Geophys. Monogr.}, \textbf{119}, 321, 2000.

Malandraki,~O.\,E., N.~Agueda, A.~Papaioannou, K.--L.~Klein, E.~Valtonen, B.~Heber, W.~Dr{\"o}ge, H.~Aurass, A.~Nindos, N.~Vilmer, B.~Sanahuja, A.~Kouloumvakos, S.~Braune, P.~Preka--Papadema, 
K.~Tziotziou, C.~Hamadache, J.~Kiener, V.~Tatischeff, E.~Riihonen, Y.~Kartavykh, R.~Rodr{\'{\i}}guez-Gas{\'e}n and R.~Vainio, Scientific Analysis within SEPServer -- New Perspectives in Solar Energetic Particle Research: The Case Study of the 13 July 2005 Event, \textsl{Sol. Phys.}, \textbf{281}, 333--352, 2012.

Magdaleni{\'c},~J., C.~Marqu{\'e}, A.\,N.~Zhukov, B.~Vr{\v s}nak and A.~Veronig, Flare-generated Type II Burst without Associated Coronal Mass Ejection, \textsl{Sol. Phys.}, \textbf{746}, 152, 2012.

Melnik,~V.~N., A.~I.~Brazhenko, A.\,A..~Konovalenko, C.~Briand, V.~V.~Dorovskyy, P.~Zarka, A.~V.~Frantsuzenko, H.~O.~Rucker, B.~P.~Rutkevych, L.~Denis, T.~Zaqarashvili and B.~Shergelashvili, Decameter Type III Bursts with Changing Frequency Drift-Rate Signs, \textsl{Sol. Phys.}, \textbf{290}, 193--203, 2015.

Nigl,~A., P.~Zarka, J.~Kuijpers, H.~Falcke, L.~B\"ahren and L.~Denis, VLBI observations of Jupiter with the initial test station of LOFAR and the Nan{\c c}ay decametric array, \textsl{Astron. Astrophys.}, \textbf{471}, 1099--1104, 2007.

Pick,~M., G.~Stenborg, P.~D{\'e}moulin, P.~Zucca and A.~Lecacheux, Homologous Solar Events on 2011 January 27: Build-up and Propagation in a Complex Coronal Environment, \textsl{Astrophys. J.}, \textbf{832}, 5, 2016.

Salas--Matamoros,~C., K.--L.~Klein and A.\,P.~Rouillard, Coronal mass ejection-related particle acceleration regions during a simple eruptive event, \textsl{Astron. and Astrophys.}, \textbf{590}, A135, 2016.

Shevchuk,~N.~V., V.~N.~Melnik, S.~Poedts, V.~V.~Dorovskyy, J.~Magdalenic, A.~A.~Konovalenko, A.~I.~Brazhenko, C.~Briand, A.~V.~Frantsuzenko, H.~O.~Rucker, and P.~Zarka, The storm of decameter spikes during the event of June 14, 2012, \textsl{Solar Physics}, \textbf{291}, 211--228, 2016.

Zarka,~P., Trois d{\'e}cennies d'{\'e}tude des {\'e}missions radio de Jupiter : du R{\'e}seau D{\'e}cam{\'e}trique de Nan{\c c}ay {\`a} LOFAR, Ecole CNRS de Goutelas, \textbf{30}, 205--215, 2007.

Zarka,~P., J.~N.~Girard, M.~Tagger and L.~Denis, LSS/NenuFAR: The LOFAR Super Station project in Nan{\c c}ay, SF2A-2012: Proceedings of the Annual meeting of the French Society of Astronomy and Astrophysics, Eds.: S. Boissier, P. de Laverny, N. Nardetto, R. Samadi, D. Valls--Gabaud and H. Wozniak, 687--694, 2012.

Zucca,~P., M.~Pick, P.~D{\'e}moulin, A.~Kerdraon, A.~Lecacheux and P.\,T.~Gallagher, Understanding Coronal Mass Ejections and Associated Shocks in the Solar Corona by Merging Multiwavelength Observations, \textsl{Astrophys. J.}, \textbf{795}, 68, 2014.

\end{document}